\documentclass[a4paper,12pt]{article}
\usepackage[dvipdfmx]{graphicx}
\usepackage{cite}
\usepackage{amsmath}
\usepackage{amsfonts}
\usepackage{amssymb}
\usepackage{graphicx, rotating}
\usepackage{epsfig}
\usepackage{latexsym}
\usepackage{graphicx}
\usepackage{color}
\usepackage{amsmath,bm,amssymb}

\usepackage{color}
\usepackage[english]{babel}
\usepackage{hyperref}

\newcommand{\Slash}[1]{{\ooalign{\hfil#1\hfil\crcr\raise.167ex\hbox{/}}}}

\newcommand{\beq}{\begin{equation}}  \newcommand{\eeq}{\end{equation}}
\newcommand{\bef}{\begin{figure}}  \newcommand{\eef}{\end{figure}}
\newcommand{\bec}{\begin{center}}  \newcommand{\eec}{\end{center}}
\newcommand{\non}{\nonumber}  
\newcommand{\laq}[1]{\label{eq:#1}}  

\newcommand{\Eq}[1]{Eq.~(\ref{eq:#1})}
\newcommand{\Eqs}[1]{Eqs.~(\ref{eq:#1})}
\newcommand{\eq}[1]{(\ref{eq:#1})}
\newcommand{\Sec}[1]{Sec.\ref{chap:#1}}

\newcommand{\vev}[1]{ \left\langle {#1} \right\rangle }

\newcommand{\lac}[1]{\label{chap:#1}}
\newcommand{\SU}[1]{{\rm SU{#1} } }

\def\({\left(}
\def\){\right)}

\def\O{\mathcal{O}}
\def\U{\mathop{\rm U}}

\newcommand{\OR}{~{\rm or}~}
\newcommand{\AND}{~{\rm and}~}
\newcommand{\EV}{ {\rm \, eV} }
\newcommand{\KEV}{ {\rm \, keV} }
\newcommand{\MEV}{ {\rm \, MeV} }
\newcommand{\GEV}{ {\rm \, GeV} }
\newcommand{\TEV}{ {\rm \, TeV} }

\def\a{\alpha}

\def\d{\delta}
\def\e{\epsilon}
\def\f{\phi}
\def\g{\gamma}

\def\k{\kappa}
\def\l{\lambda}
\def\m{\mu}
\def\n{\nu}
\def\p{\psi}

\def\s{\sigma}
\def\t{\tau}

\def\D{\Delta}
\def\G{\Gamma}
\def\L{\Lambda}
\def\F{\Phi}

\def\tl{\tilde}
\def\*{\dagger}

\begin{document}
\renewcommand\bibname{\Large References}

\begin{flushright}
TU-1138
\end{flushright}

\begin{center}

\vspace{1.5cm}

{\Large\bf Phenomenology of CP-even ALP}\\
\vspace{1.5cm}
\renewcommand{\thefootnote}{\fnsymbol{footnote}}
{\bf  Kodai Sakurai$^{1}$
\footnote{E-mail:  \texttt{kodai.sakurai.e3@tohoku.ac.jp}}}
and {\bf Wen Yin$^{1}$
\footnote{E-mail:  \texttt{yin.wen.b3@tohoku.ac.jp}}}

\vspace{12pt}
\vspace{1.5cm}
{\em 

$^{1}${Department of Physics, Tohoku University, Sendai, Miyagi 980-8578, Japan} 
}

\vspace{1.5cm}
\abstract{
Axion or axion-like particle (ALP) has been usually considered as a CP-odd Nambu-Goldstone boson  (NGB) from the spontaneous breakdown of a global U(1) symmetry. 
In this paper, we point out that 
the NGB behaves as a CP-even particle coupled to the SM particles in a large class of simple (or perhaps the simplest) renormalizable models. We provide a first study of the collider phenomenology and cosmology of the CP-even ALP. 
In a natural parameter region, the CP-even ALP can be produced from the Higgs boson decay in colliders. 
When the mass is not very light, the signals will be Higgs exotic decays, Higgs decay to displaced vertex $\times 2$, Higgs decay to displaced vertex + missing energy. 
The signal can be discriminated from other models, e.g. hidden photon, by measuring the decay length and the decay products of the light new particle. 
In addition, when $ m_a\lesssim \MEV$, in which case the Higgs boson invisible decay may be probed in the colliders, 
the CP-even ALP  is a nice Dark matter (DM) candidate. 
The DM can be probed by 21cm line measurement, the future measurement of the Milky way mass halo function in the Vera Rubin Observatory, as well as X- or $\gamma$-ray observations.
The DM production mechanisms are discussed.
}

\end{center}
\clearpage

\setcounter{page}{1}
\setcounter{footnote}{0}


\section{Introduction}

The existence of a dark sector is plausible due to the evidence of dark matter (DM). 
In particular, a light dark sector may be reasonable since DM stability is easily guaranteed. 
The axion-like particle (ALP) coupled to the standard model (SM) particles is a leading candidate of the DM, and they arise from the spontaneous breaking of a global U(1) symmetry\footnote{It is called as Peccei-Quinn symmetry when the U(1) is anomalous to some gauge symmetry~\cite{Peccei:1977hh,Peccei:1977ur,Weinberg:1977ma,Wilczek:1977pj}.} whose Nambu-Goldstone boson (NGB) is the ALP (see
Refs.\,\cite{Jaeckel:2010ni,Ringwald:2012hr,Arias:2012az,Graham:2015ouw,Marsh:2015xka,Irastorza:2018dyq,DiLuzio:2020wdo} for reviews.). 
The lightness of the axion/ALP is because of the smallness of the explicit breaking term of the symmetry. 
It is usually assumed that the ALP potential or the interaction is CP-conserving and the ALP is a CP-odd particle 
despite that the CP-symmetry is absent in the SM and despite that CP-conserving nature of the QCD is dubbed as a strong CP problem.

In this paper, we remove the assumption of the CP-symmetry of the dark sector for the  ALP,\footnote{This assumption was once removed in the following studies. 
In Ref.~\cite{Marsh:2019bjr}, a model with a QCD axion and a string ALP of mass $\sim 10^{-15}\EV$ was studied from motivation by an M-theory compactification. 
It was found that the mixing between the QCD axion and ALP can induce a sufficient CP violation which may be measured in future EDM experiments. 
In Ref.~\cite{Kim:2021eye}, a long-range force between the CP violating dark sector and the CP conserving SM sector was studied. It was found that the axion force can be 
measured from the daily modulation of the matter spin precession. Interestingly the effective magnetic field is towards to the galactic center.} and point out that in a simple (or perhaps the simplest) renormalizable model without any other BSM fields than the dark Higgs field, 
 the low energy effective theory (EFT) has an accidental CP symmetry, under which 
the ALP is CP-even.
The couplings involving single {\it CP-even ALP} to the SM particles exist due to the mixing with the SM Higgs boson, and they are suppressed by the ALP mass squared, $m_a^2$. 
On the contrary, the very weakly coupled ALP at the low energy scale is not necessarily weakly coupled to the Higgs boson. 
By taking this property we provide a first study of the collider phenomenology and cosmology of this CP-even ALP. 
In fact, future Higgs factories such as HL-LHC and ILC
provide a great opportunity to search for the CP-even ALP from the Higgs boson decay. 
The typical signatures are a SM Higgs boson exotic decay, displaced vertices/displaced vertex plus missing energy followed by the Higgs boson decay or a Higgs invisible decay depending on the decay length. Measuring the decay product and decay length provides distinguishable signature of the model, e.g. from hidden photon which is discussed in the appendix. 
The decay length (at rest) is longer than the age of Universe when the mass is smaller than $\MEV$ thanks to the highly suppressed decay rate $\propto m_a^{7}.$
The ALP DM can be probed from the future observations of X, $\g$-ray as well as the 21 cm line if the ALP is produced thermally. 
The DM production mechanisms are also discussed.

Let us refer to several studies and the difference of this work. 
A NGB WIMP DM was discussed in e.g. Refs.\,\cite{Ishiwata:2018sdi, Cline:2019okt, Abe:2020iph}, 
where the authors assumed that the $\U(1)$ symmetry is explicitly broken due to a dimension-2 parameter preserving a parity symmetry and assumed a portal coupling between the dark Higgs and SM Higgs bosons. 
In this case, the ALP potential is symmetric under an accidental C-symmetry only in the dark sector, and the C-symmetry guarantees the stability of the ALP. 
On the other hand, the highly suppressed ALP-SM photon coupling may be similar to the anomaly-free or photophobic ALP model~\cite{Pospelov:2008jk,Arias:2012az, Nakayama:2014cza, Takahashi:2020bpq, Takahashi:2020uio}.
In any case, the ALPs can be defined as CP-odd, which is the clear difference from our proposal of CP-even ALP.

\section{Models of CP-even ALP} 

\subsection{A renormalizable model of ALP and SM Higgs boson decay}
\lac{1}

A minimal light dark sector for our purpose has one dark Higgs field, $\F$,  which spontaneously breaks the hidden global $\U(1)$ symmetry by its vacuum expectation value (VEV), 
\beq 
\vev{\F}=v_\F. 
\eeq
In this minimal setup, the only renormalizable interaction between the SM and dark sector is the portal coupling between the SM and dark Higgs field.
The most general dark and SM Higgs potential is given by\footnote{The same potential was used to study the light ALP/hidden photon DM production via a second order phase transition in Ref.\,\cite{Nakayama:2021avl}. In this scenario, the portal coupling should be so small that the hidden sector is not thermalized in the early Universe. }
\beq
V=-m^2_\F|\F|^2+{\lambda} |\F|^4 +\l_P |H|^2 |\F|^2 + \lambda_H |H|^4-\mu_H^2 |H|^2. \label{V}
\eeq 
Here $\F$ ($H$) is the hidden (SM) Higgs field (doublet) which will break the $\U(1)$ ($\SU(2)_L\times \U(1)_Y$) symmetry,  
$\lambda_P, \l (>0) \AND \l_H (>0)$ are coupling constants,  $\m_H^2 \simeq (125\GEV)^2/2$ is the bare Higgs mass term in the SM, and $m^2_\F (>0)$ is the dark Higgs mass squared parameter. 
Here $v_\F \approx \sqrt{\frac{m_\F^2}{2\lambda }}$.
For an absence of tuning for both the SM Higgs boson mass and dark Higgs mass we need 
\beq
|\l_P| v^2 \lesssim m_\F^2,|\l_P| v_\F^2 \lesssim \m_H^2. 
\eeq
Here $v=\vev{H}\approx 174\GEV$ is the Higgs field VEV. 
Thus, we cannot have a large hierarchy between the SM Higgs and dark Higgs masses if $\l_P$ is not tuned to be small. 
It may be noteworthy that there is no symmetry to be recovered at $\l_P\to 0$ and thus small $\l_P$ is not natural~\cite{tHooft:1979rat}. 
The SM Lagrangian with this potential, or more generically with the form of $V=V(|\F|^2, |H|^2)$, has two accidental discrete symmetries: 
\begin{description}
\item[$C_{\rm dark}$ symmetry:] SM fields do not transform, $\F(t,\vec{x})\to \F^*(t,\vec{x})$ 
\item[$CP$ symmetry:] SM fields transform as in the SM, $\F(t,\vec{x})\to  \F^*(t,-\vec{x}).$
\end{description}
The CP symmetry conserves if the contributions from the phases of the CKM matrix and strong CP are not important. In this paper, we do not consider the effect of the CKM and strong CP phases.

After the symmetry breaking we obtain 
\beq
\F = (v_\F +s/\sqrt{2})\exp{( i a/f_a)}
\eeq
where $a$ is the NGB with the decay constant $f_a=\sqrt{2} v_\F$ and $s$ is the dark Higgs boson. 
The NGB will acquire a mass term via an explicit breaking term of $\U(1)$.

Now let us suppose $m_s >m_{h}/2$, 
where 
the masses for $s/h$ are derived as
\beq 
m_{s/h}^{2}=2\left(\lambda_{H}v^{2}+\lambda_{\Phi}v_{\Phi}^{2}\pm\sqrt{\lambda^{2}_{H}v^{4}+\lambda^{2}_{\Phi}v_{\Phi}^{4}+v^{2}v_{\Phi}^{2}(\lambda_{P}^{2}-2\lambda_{H}\lambda_{\Phi}) }\right),
\eeq
using the tadpole conditions for $H$ and $\Phi$, $\mu_{H}^{2}=2\lambda_{H}v^{2}+\lambda_{P}v_{\Phi}^{2}$ and $m_{\Phi}^{2}=\lambda_{P}v^{2}+2\lambda_{\Phi}v_{\Phi}^{2}$. 
In this case, the dark Higgs boson  $s$ is not be produced via the SM Higgs boson decay. 
Then, we can discuss the SM Higgs physics in an effective theory by integrating out $s$
  \beq
  \laq{effint}
{\cal L}_{\rm eff} = \frac{1}{\L_H^2}  (\partial a)^2 |H|^2\to \frac{\sqrt{2} v}{\L_H^2} h  (\partial a)^2,
  \eeq
where
\beq
\frac{1}{\L_H^2}\equiv - \frac{\l_P  }{m_s^2-m_h^2}.
\eeq
In the low energy effective theory, only this term connects the NGB with the Higgs boson or some light SM particles by further integrating out the Higgs boson.
In particular, the Higgs boson can decay into the NGB pair at the rate 
\beq
\laq{hinv}
\G_{h \to aa }\simeq   \frac{1}{16\pi} \frac{ v^2 m_h^3}{\Lambda_H^4}.
\eeq
by neglecting the NGB mass. 
This contributes to the branching fraction of the Higgs boson decay as 
\beq
{{\rm Br}_{h\to aa}= {1.9\%} \(\frac{2\TEV}{\L_H}\)^4}.
\eeq

As we will see shortly, 
by explicit breaking of the $\U(1)$ without imposing CP symmetry, the NGB can decay into the SM particles depending on the range of the masses and couplings. 
This will provide detectable signature of the light dark sector in colliders and in 
DM indirect detection experiments.

\subsection{CP-even ALP and colliders }\label{sec:col}

Now we introduce a mass term to the NGB (or ALP) by explicitly breaking the $\U(1)$ global symmetry. 
The generic renormalizable terms are given by 
\beq\label{Vexp}
\d V=  \kappa \(\sum_{j=1}^4 c_j  m_\F^{4-j} \Phi^{j}+ \sum_{j=1}^2 (\tl c_j^H m_\F^{2-j} \Phi^{j}  |H|^2+\tl c_j^\F m_\F^{2-j} \Phi^{j}  |\F|^2) \) +{\rm h.c.}
\eeq
where $\k$ is a real order parameter  for the $\U(1)$ explicit breaking, i.e. at $\k\to 0$ the $\U(1)$ symmetry is exact. 
Thus $\k$ can be naturally small~\cite{tHooft:1979rat}. 
$c_j, \tl c^x_j$ $( x= H, \Phi)$ are complex dimensionless coefficients, the size of which is $\O(1)$.

By integrating out $h$ and $s$ we obtain 
\begin{align}
\laq{Va}
&V_a= \non \\
 -\kappa  &\( \sum_{j=1}^{4} |c_j| v_\Phi^j m_\F^{4-j} \cos(j \frac{a}{f_a} +\theta_j )+
\sum_{j=1}^2 |\tl c^H_j| v_\Phi^j v^2 m_\F^{2-j} \cos(j \frac{a}{f_a} +\tl \theta^H_j )+
 |\tl c^\F_j| v_\Phi^j v^2 m_\F^{2-j} \cos(j \frac{a}{f_a} +\tl \theta^\F_j )\)
\end{align}
where $\theta_j = \arg c_j, \tl \theta^x_j= \arg \tl c^x_j$.
This immediately provides the mass to the NG boson around the minima of $V_a$:
\beq
 m_a^2 \sim \O(\k) {v_\F^2}\sim \O(\k) m_s^2
\eeq
where we have assumed $v_\F\sim m_s\gtrsim v.$  
The small mass is natural because at the vanishing limit of $\k$ the mass vanishes. 

In the generic Lagrangian not only the mass of the ALP is obtained, but also an explicit $C_{\rm dark}$-violation can be obtained in the case with generic $\theta_j, \tl\theta_j^{x} \neq 0$.  Thus $\vev{a}\neq 0$, and  $a$  mixes with $s$ and $h$. 
Note that if $\vev{a}\neq 0 $, stabilizing the potential by $a$ is a cancellation among the first derivatives of several cosine terms.
On the other hand, $a-h$ or $a-s$ mixing needs not be canceled. 
This can be found explicitly from the following replacement of the parameters in \Eq{Va}: 
\beq v_\F\to v_\F (1+ \frac{s}{\sqrt{2}v_\F}) \AND v\to v (1+\frac{h}{\sqrt{2}v}).
\eeq
The mixing between $a$ and $s (h)$ can be obtained by taking { the} 
derivatives with respect of  $a$ and $s (h)$ around $a=\vev{a}$. Although at around $a=\vev{a}$ the first derivatives of $a$ in \Eq{Va}  is vanishing, after the recovery of $s$ and $h$, it is generically non-vanishing since the powers of $ (1+ \frac{s}{\sqrt{2}v_\F}) \OR (1+\frac{h}{\sqrt{2}v})$ are different for different $j$ of the cosines.

There are two types of the $a$-$h$ mixings. 
The first one is from the product of the mixings between $a\text{-}s$ and between $s$-$h$. The $a$-$s$ mixing is estimated by (see the first and last terms of \Eq{Va})
$
\theta_{as}\sim  \k v_\F m_\F /m_s^2
$
while $s$ - $h$ mixing is not suppressed by $\k$, $\theta_{sh}\sim \l_P {v} v_\F/m_s^2$ where we have assumed $m_s \gg m_h$ and $c_{j}, \tl c^{x}_{j}$ {(also the phases $\theta_{j}$, $\tl \theta_{j}$)} are $\O(1)$ for simplicity. 
In total we get the mixing of order $\theta_{a h}^{(\F)}\sim \l_P \k \frac{v_\F^2 m_\F v }{m_s^4}$. 
By using ${v} \sim m_h, v_\F \sim m_s \sim m_\F,$ $\k \sim m_a^2/m_\F^2$ and $\l_P\sim \O(1)$, we obtain $\theta_{a h}^{(\F)}\sim \frac{ m_a^2 v}{m_s^3}.$ 

The other contribution is the direct $h$ - $a$ mixing which naturally arises by including the middle terms of \Eq{Va}.  
This is given by $\theta_{ah}^{H}\sim \k {v} v_\F/m_h^2\sim m_a^2/m_\F m_h.$
{The direct $h$ - $a$ mixing }$\theta_{ah}^{H}$ has a  size larger than $|\theta_{ah}^\F|$ when $m_s \gtrsim m_h$ with $\O(1)$ parameters (except for small $\k$).
Thus, we parameterize the mixing in a model-independent way 
\beq
\laq{Defc_h}
{\theta_{ah}= c_h \frac{m_a^2}{m_h m_\F} },
\eeq
with $c_h$ being a dimensionless model dependent parameter, which is a  function of the parameters in the original Lagrangian.  

\begin{figure}[!t]
\begin{center}  
   \includegraphics[width=.45\textwidth]{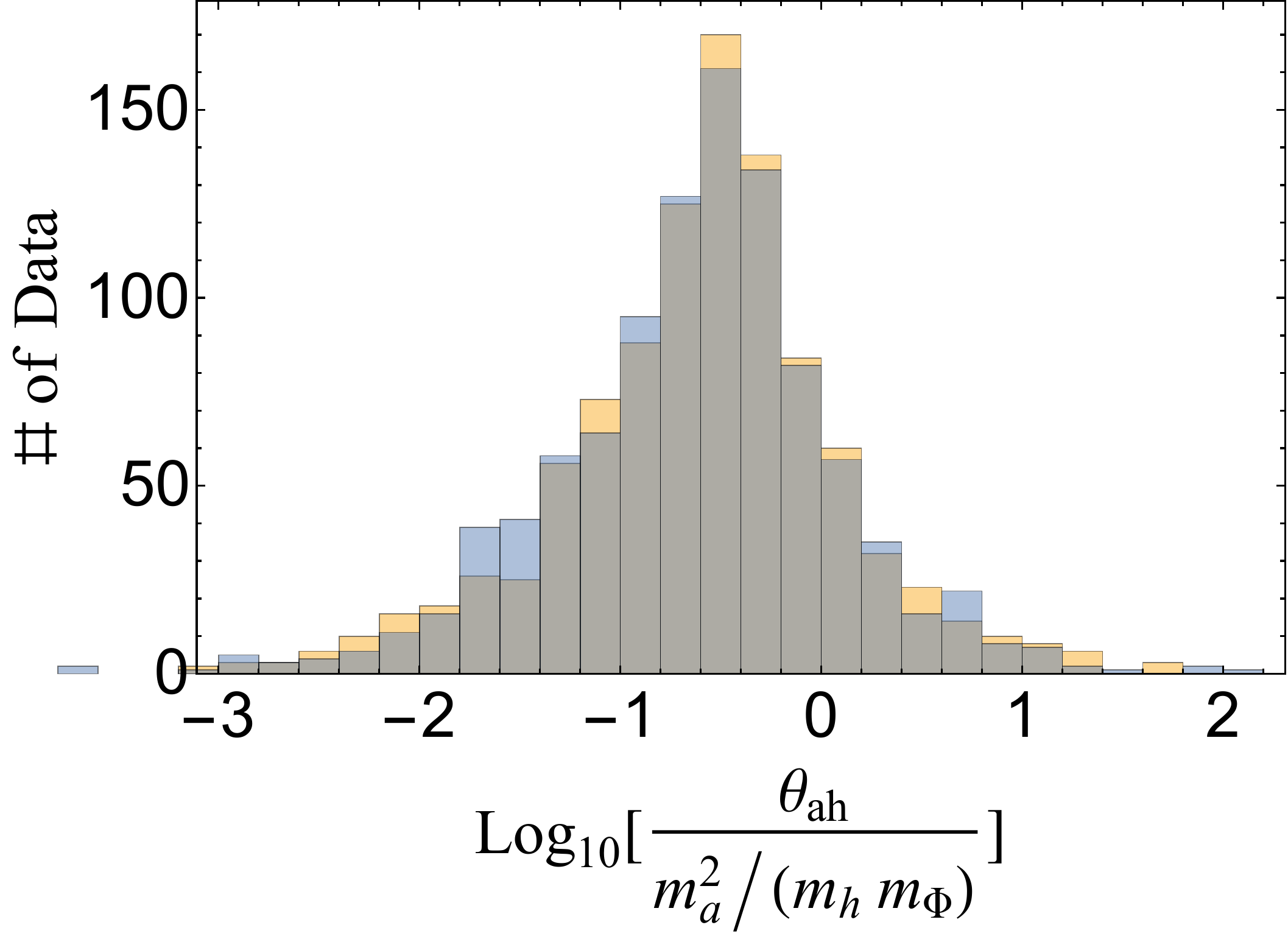}\hspace{1.5mm}
   \includegraphics[width=.45\textwidth]{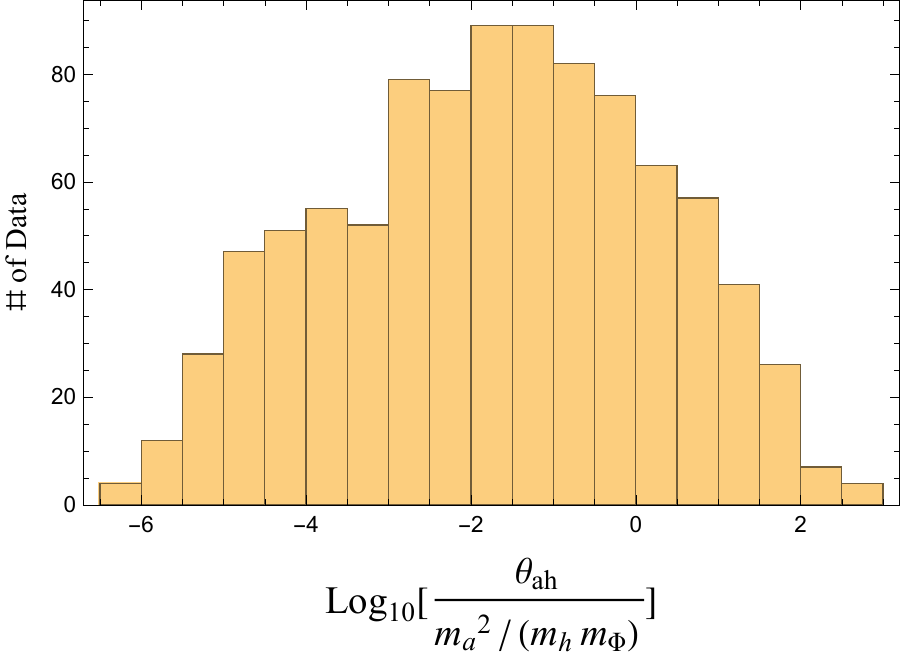}
      \end{center}
\caption{The histogram of $|\theta_{ah}|/(m_a^2/(m_h m_\F))$ in  scatter plot with effective theory analysis (left panel) and full theory analysis (right panel).   
In the left panel, the data of the orange and blue region represent that $|\theta_{ah}|/(m_a^2/(m_h m_\F))$ is evaluated in true and false vacua, respectively. They overlap in gray. 
In the right panel, we do not specify whether the vacuum is false or true. 
 } \label{fig:1} 
\end{figure}

This relation is checked numerically in two kinds of parameter scans. 
In the left panel of Fig.~\ref{fig:1} we take 1000+1000 points randomly in the following range: 
$|c_{1,2}|, |\tl{c}^{H}_{1}|=[0-1]$ with arbitary phases; other $c$s are taken to be zero i.e. there can be two vacua for each of which we provided 1000 points; $m_s=(500-10^4)\GEV$, $v_\f=(1-100)m_s$ and 
$\l_P<0.1$.
We stabilize the potential of $a$ in the effective theory of $V_a$. There are at most two vacua. The data in the false and true vacuum are shown in  blue and orange data, respectively. Thus there are no significant differences. In the right panel, we perform a numerical analysis of the full theory including the whole potential of $\F \AND H$. The strategy of the parameter scan is shown in appendix \ref{ap:mass}, where we do not specify whether the vacuum is true or not.

Through this mixing, the ALP couples to the SM particles and can decay into them. 
The decay rate of $a$ to a set of SM particles $\O$ can be evaluated from the SM-like Higgs boson decays as  
\beq\label{eq:adecay}
\G_{a \to \O }[m_a] = \theta_{ah}^2 \G_{h \to \O }|_{m_h \to m_a}= c_h^2\frac{ m_a^4}{m_h^4} [\G_{h\to \O}]_{m_h\to m_a}.
\eeq
The decay rates of the SM-like Higgs boson are taken from ~Refs.~\cite{Djouadi:2005gi,Spira:2016ztx}. 
For the decay into hadrons, we use the results given in Ref.~\cite{Winkler:2018qyg}.

Interestingly, the ALP looks like a CP-even scalar from its coupling to the SM. The induced EDMs are highly suppressed. 
This can be understood from the symmetry.
 As we have mentioned, at $\k\to 0$ the action is invariant under both  $C_{\rm dark} \AND CP$ symmetry. 
In the presence of $\k\neq 0$  and generic $\theta, \tl\theta^x \neq 0$, both $C_{\rm dark}$ and $CP$ are explicitly broken. 
On the contrary \begin{description}
\item{ \boxed{CP_{\rm EFT}\equiv C_{\rm dark}\cdot CP}:} the SM fields transform as in the SM, $\Phi[t,\vec{x}]\to \Phi[t,-\vec{x}]$ 
\end{description}
remains as a good symmetry.\footnote{Indeed, our model is minimal flavor and CP violating. It is flavor or CP safe. That said, if we introduce CP violating ALP coupling in addition to $C_{\rm dark}$ breaking, the model is then  constrained by the EDMs.}
Although $CP_{\rm EFT}$ is not the usual CP symmetry, 
in the low energy effective theory by integrating out $s$, it looks like a CP symmetry\footnote{This is the reason that we call $CP_{\rm EFT}$ just as CP in the effective theory.}, under which the ALP $a$ is CP even. Therefore we predict a {\it CP-even ALP}.

An interesting observation of this model, in contrast to the usual ALP model,  
is that the interaction between the Higgs and an ALP is not suppressed but the interaction between the ALP and the light SM particles is (highly) suppressed by the positive powers of the tiny ALP  mass. 
This property is useful in probing it in colliders although the ALP is weakly coupled to low energy physics. 
To clarify this, let us estimate the decay length of the ALP defined by   
\beq
{L_D}\equiv \G^{-1}_{\rm tot} \frac{E_{a}}{m_{a}},
\eeq 
where $E_{a}$ is the energy of $a$ in the Laboratory frame. 
$E_a\sim m_h/2$ if $a$ is produced in $E_{\rm cm}= 250\GEV$ lepton colliders, e.g. $E_h\sim 140\GEV$ via $e \bar e \to Z h$, and $E_a \sim 40-100\GEV.$ 
Depending on the relative relation among  detector volume, $L_V$,  detector resolution $L_R (< L_V)$, and the decay length, $L_D$, we have different collider signatures. 
The decay happens within the detector at the probability of $1-e^{-L_V /L_D}$ and can be recognized as the displaced vertex at the probability of
$ e^{-L_R/L_D}.$ Thus a displaced vertex for a single $a$ decay happens at the probability of $e^{-L_R/L_D}(1-e^{-L_V /L_D})$ if the decay products can be measured. 
Therefore, $a$ may or may not decay within the detector, and may or may not be recognized as the displaced  vertex depending on $L_D$ given the 
detector properties of $L_R$ and $L_V$.
The typical signals depend on the relative size of $L_D, L_V, L_R$ which is shown in Table. \ref{tab:1}.  

One interesting thing is that we can measure $L_D$ in some cases. 
For instance, $L_D$ is around the most distant displaced vertices in the regime of $L_V>L_D>L_R$. 
$L_D\sim 2 \frac{\#_\text{Invisible decays}}{\#_\text{Displaced + missing} } L_V \sim 1/2 \frac{\#_\text{Displaced + missing} }{\#_\text{Displaced $\times 2$} } L_V$  in the regime $L_D \sim L_V>L_R$.\footnote{In the case Higgs boson also invisibly decay to other BSM particles the first form in the equation will be incorrect. 
We may confirm this estimation by checking both relations if we can have sufficient number of the events. 
In addition, we can further probe the CP even ALP via beam dump experiment e.g. \cite{Blumlein:2013cua, Ilten:2018crw, Bauer:2018onh, Kanemura:2015cxa, Sakaki:2020mqb, Asai:2021xtg}, which do not require the NGB coupling to the Higgs boson. The collision center-of-mass energy for existing or near-future beam dump experiments is much smaller than $m_h$, 
and the decay volume is of the order of $\O$(10-100)m.  
In case in the Higgs factory, one has signals of large $L_D$, 
we obtain a strong motivation to have a beam dump experiment to further search for the ALP via the coupling responsible for its decay. A vice versa approach is possible to search for the Higgs-ALP coupling. 
Our approach and the beam dump approach complements each other. }

\begin{table}[htp]
\caption{Event signature in the Higgs factory with respective to  the relative size among the detector volume, $L_V$, detector resolution, $L_R$, and the decay length, $L_D$.}
\begin{center}
\begin{tabular}{|c|c|}
\hline
Condition & Signals in Higgs factory\\\hline 
$L_D \gg L_V> L_R$ & Higgs invisible decay\\\hline
 &Displaced vertex $\times 2$,  Higgs invisible decay\\
$L_D \sim L_V> L_R $ &  or/and  Displaced vertex+missing energy \\\hline 
$ L_V > L_D > L_R $  &Displaced vertex $\times 2$ \\ \hline
$ L_V >  L_R\gtrsim L_D  $  & Exotic Higgs  decay \\ \hline
\end{tabular}
\end{center}
\label{tab:1}
\end{table}%

\begin{figure}[!t]
\begin{center}  
   \includegraphics[width=1\textwidth]{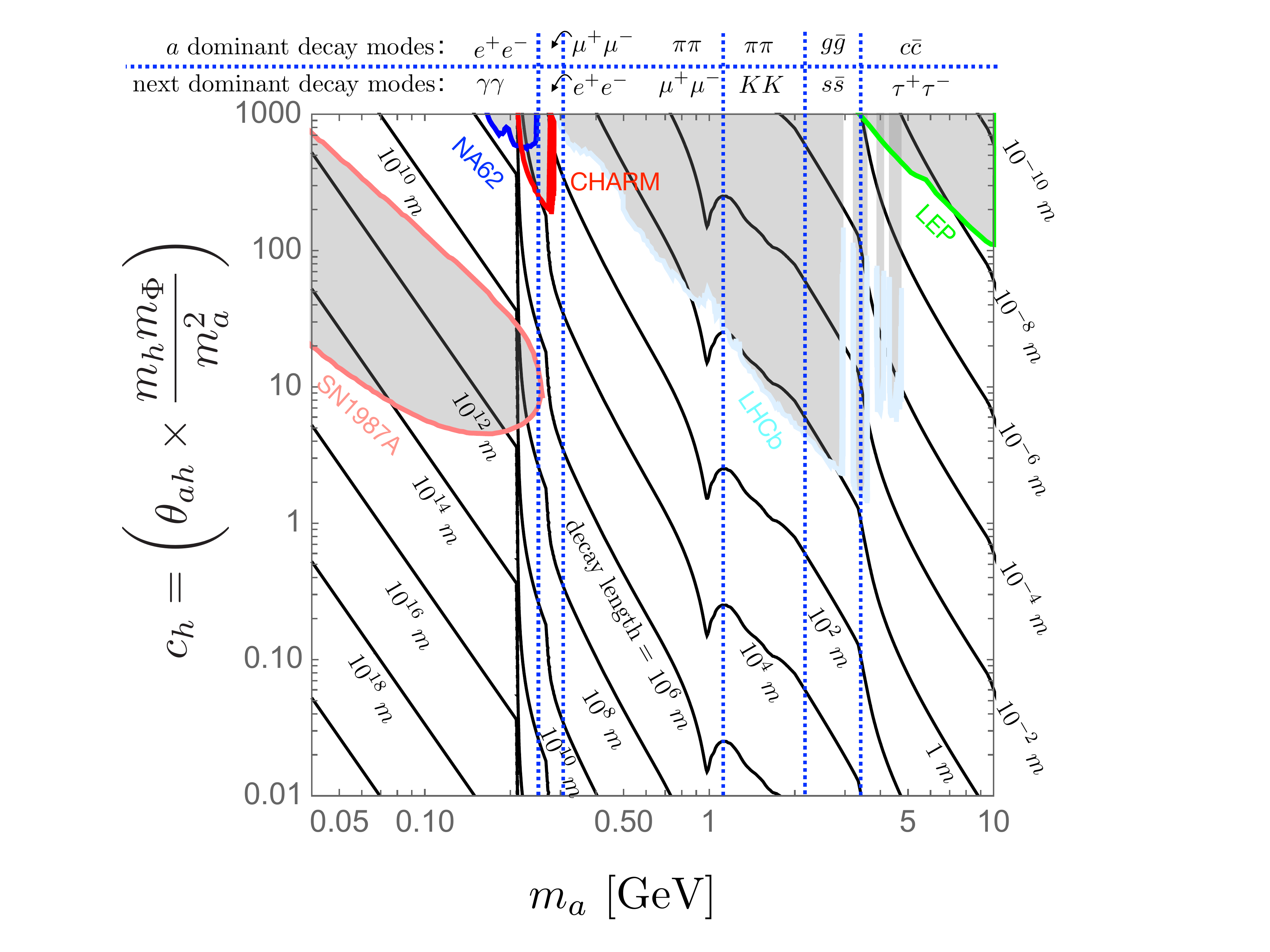}
      \end{center}
\caption{The contours of the decay length of the $a$ produced from $h\to a a$ in $m_a-c_h$ plane. 
Here we assume the initial Higgs is at rest, which is in good approximation for a $240\text{-}250\GEV$ lepton collider. 
We also take $m_{\Phi}=1{\rm TeV}$. 
 Gray regions are excluded by SN1987A~\cite{Evans:2017kti,Krnjaic:2015mbs} (pink curve), NA62~\cite{NA62:2021zjw,NA62:2020pwi,NA62:2020xlg} (blue curve), CHARM~\cite{CHARM:1985anb} (red curve),  
LHCb~\cite{LHCb:2012juf} (light blue curve) and LEP~\cite{L3:1996ome} (green curve). 
 } \label{v}
\end{figure}

Now we are ready to show the signature of dark sector of the CP-even ALP in future Higgs factories~\cite{Asner:2013psa,dEnterria:2016sca,Abramowicz:2016zbo,Apollinari:2017lan, CEPC-SPPCStudyGroup:2015csa, Ankenbrandt:1999cta, Delahaye:2019omf, Garcia:2020xrp }. For concreteness let us focus on the ILC with $250\GEV$ center-of-mass energy  and the Higgs is produced at rest (which is approximately the case if we do not consider the $\O(10\%)$ boost factor.). 
In Fig.\ref{v}, we display the contours of the decay length of $a$ from the decay of Higgs boson.
On the top of the figure, we display the dominant and next dominant decay modes. The branching ratios are evaluated from Eq.~\eqref{eq:adecay}.
Various experimental bounds are imposed in the $c_{h}$-$m_{a}$ plane, and the excluded regions are displayed in gray surrounded by colored lines. 
The blue curve corresponds to the bounds on  $K\to \pi^{+}+{\rm inv.}$ in the NA 62 experiment~\cite{NA62:2021zjw,NA62:2020pwi,NA62:2020xlg}.
The red curve comes from the CHARM beam dump experiment~\cite{CHARM:1985anb}.
The light blue curve is for the B meson decay $B^{+}\to K^{+} \mu^{+}\mu^{-}$ from the LHCb experiment~\cite{LHCb:2012juf}.
The green curve is given by the null observation of  $e^{+}e^{-}\to Z^{\ast}a$ in the L3 experiment~\cite{L3:1996ome}. 
 We recast the sensitivity of Ref~\cite{NA62:2020xlg} for the NA 62 experiment. The other constraints are taken from Ref.~\cite{Winkler:2018qyg}. 

The contours give the information for typical events. For instance, ILC may identify the decay vertex of $a$ at a distance from the Higgs decay point in $L^{\rm SID}_{\rm R}\sim 6\m {\rm m}$, $L^{\rm SID}_{\rm V} =604.2$cm \cite{Behnke:2013lya,Antusch:2016vyf} with SID and $L^{\rm ILD}_{\rm R}= 2\text{-}6\m$m, $L^{\rm ILD}_{\rm V}= 775.5$cm with ILD~\cite{Behnke:2013lya}. Here we have taken $L_{\rm R}$ as the resolution of the vertex detector, and $L_{\rm V}$ the outer radius of the muon detector. 
More conservatively we can take the $L_{\rm V}$ as the outer radius of the particle tracker or time projection chamber (TPC). 
TPC in the ILD has the outer radius of $180.8$cm.
In TPC, the trajectory and the energy loss of the charged SM particles can be measured precisely and the particle species can be identified. 
When $a$ decays inside the TPC to charged particles, there may even be no background.

When the decay length is in between $L_{V}$ and $ N_{H} Br_{H\to aa}\times L_V$, we can have both events of the displaced vertex plus missing, 
and the Higgs invisible decays. 
Here $N_H$ is the number of the Higgs boson that is produced. 
That means the ILC of 250GeV, who has the Higgs production cross-section of $\sim 300$fb~{\cite{Krnjaic:2015mbs}}, and thus, $N_H\sim 10^6$ for $3\rm ab^{-1}$ integrated luminosity can probe the displaced vertex with 
\beq
L_{D}\lesssim 10 {\rm km }\(\frac{N_H }{10^6}\) \(\frac{L_V}{2m}\) \(\frac{{\rm Br}_{h\to aa }}{1\%} \) \text{~~~($\gtrsim 2\s $CL, $L_D\gtrsim L_V$)},
\eeq 
where we have assumed the absence of the background events, and used the definition of the significance=$(\rm \# ~of~ events)^{1/2}$.

The invisible decay region can be tested in the future Higgs factories with~\cite{Asner:2013psa,dEnterria:2016sca,Abramowicz:2016zbo, CEPC-SPPCStudyGroup:2015csa, Ankenbrandt:1999cta, Delahaye:2019omf, Garcia:2020xrp, Kato:2020pyl } 
\beq
\laq{br}
Br_{h\to aa} >\O(0.01-0.1)\%  \text{~~~($L_D\gg L_V$)}.
\eeq
For ILC it is considered as $Br_{h\to aa}>\O(0.1)\%$. 
When the signal includes displaced vertex or exotic decay~(see Ref.\,\cite{Liu:2016zki}), especially for the displaced vertex, there may even no background event. 
Then the reach of the branching fraction (with the decay within the detector, especially in the ILD)
\beq\label{brex}
Br_{h\to aa} \gtrsim  2\times 10^{-4}\% \(\frac{10^6}{N_H}\)  \text{~~~($\gtrsim 2\s$ CL, $L_V>L_D>L_R$)}. 
\eeq
In the region with \Eq{br} all the region can be tested. 

\begin{figure}[!t]
\centering
   \includegraphics[width=0.45\textwidth]{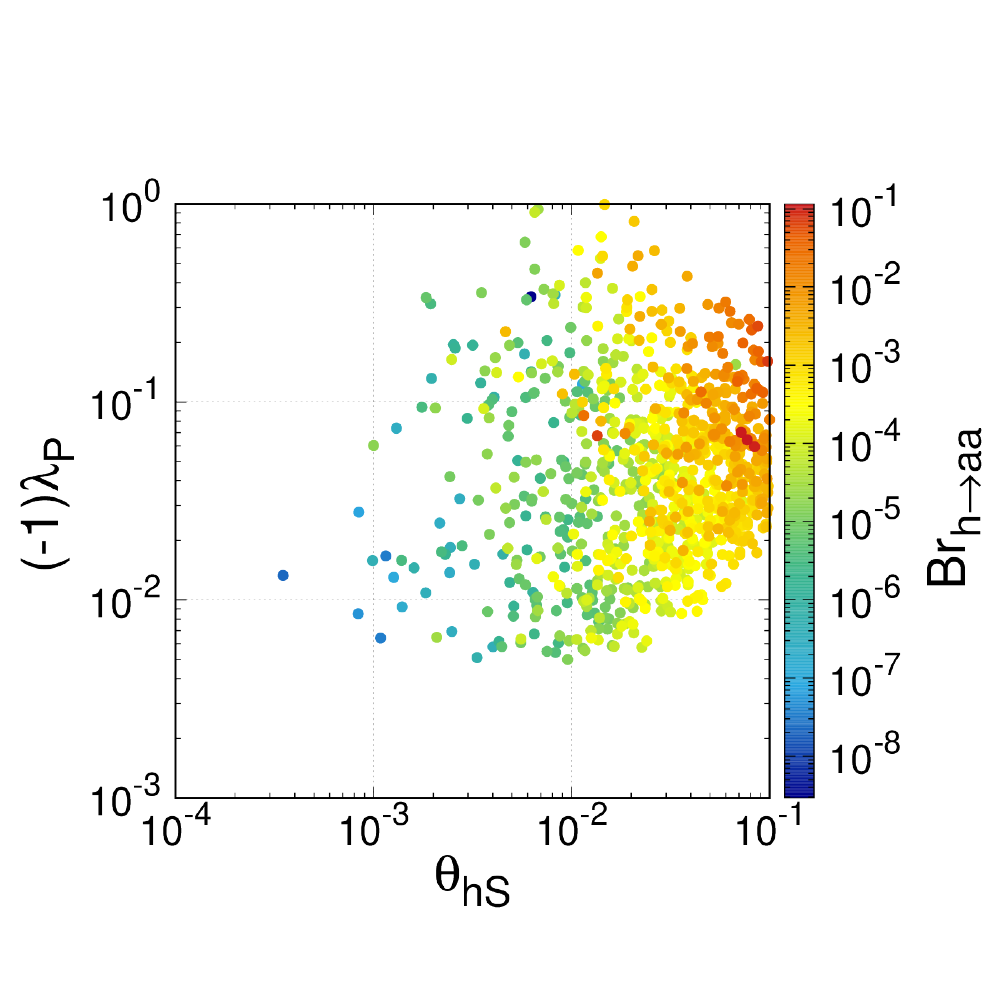} \hspace{1.5mm}
   \includegraphics[width=0.45\textwidth]{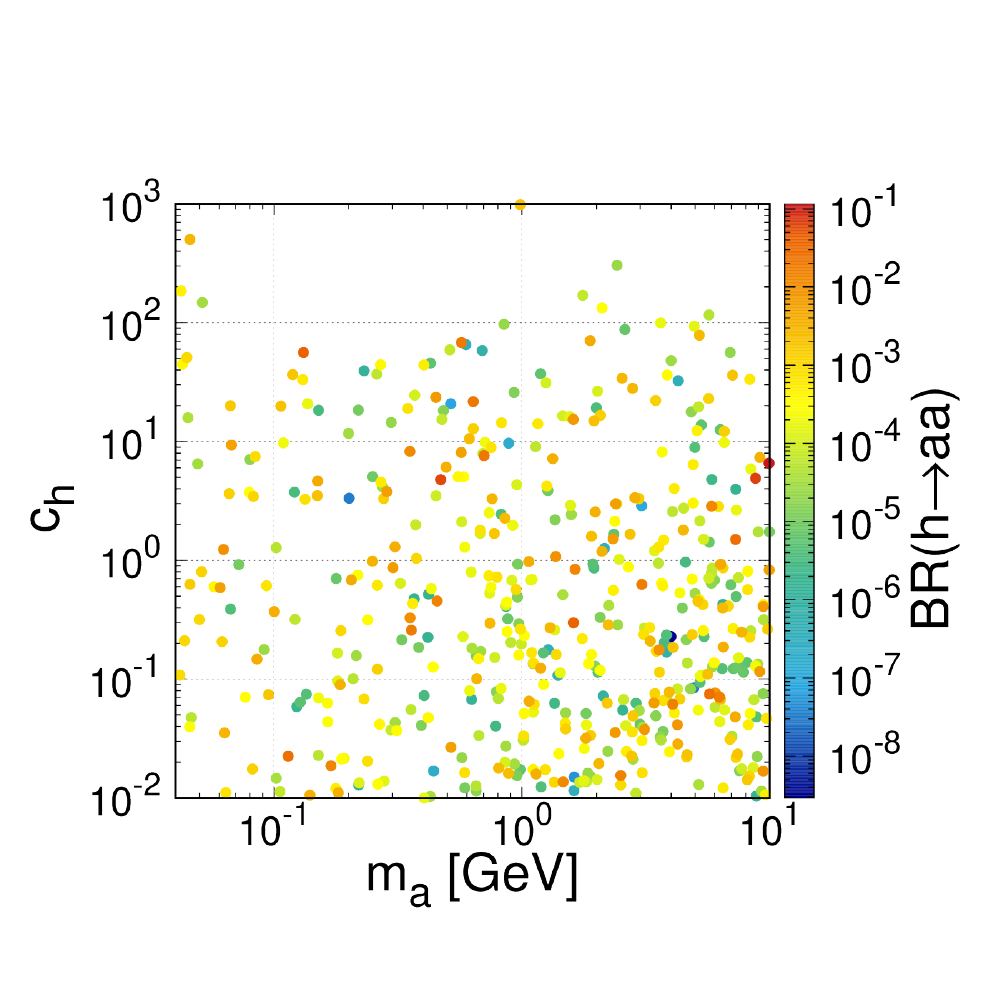}
\caption{The branching ratio for the SM Higgs boson decay into ALP, $h\to aa$. The left panel is the result in the plane of the mixing angle between $h$ - $s$ and the portal coupling $\lambda_{P}$. The right panel is the result in the plane of the mass of ALP $m_a$ and the dimensionless parameter $c_h$.  The magnitude of $Br_{h\to aa}$ is shown by the color bar. 
 } \label{fig:br} 
\end{figure}

In Fig.~\ref{fig:br}, we show the branching ratio of the SM Higgs boson into ALP in the plane of the portal coupling $\lambda_{P}$ and the $h$ - $a$ mixing $\theta_{sh}$ as well as the mass of ALP $m_a$ and the dimensionless parameter $c_h$.
In order to evaluate the branching ratio, we use the mass basis for the Higgs bosons (see the detail in the Appendix~\ref{ap:mass}).  The branching ratios are evaluated by 
\begin{align}
{\rm Br}_{h\to aa} =\frac{\Gamma_{h\to aa}}{\Gamma_{h\to SM}+\Gamma_{h\to aa}}, 
\end{align}
where $ \Gamma_{h\to SM}$ denotes the decay rate for the SM-like Higgs boson decay into SM particles.  
It is expressed in terms of the mixing angles and the SM prediction as $\Gamma_{h\to SM}=(\cos\alpha_{1}\cos\alpha_{2})^{2}\Gamma_{h\to SM}^{\rm SM}$. 
The SM prediction for Higgs boson decay into SM particles is given in Refs.~\cite{Djouadi:2005gi,Spira:2016ztx}. For the decay into a pair of quarks, two gluons, two photons as well as a photon and a Z boson, we include the NLO QCD corrections. 
The decay rate for $h\to aa$ is written by
\begin{align}
\Gamma_{h\to aa}=\frac{(2\lambda_{haa})^{2}}{32\pi m_{a}}\sqrt{1-4\frac{m_{a}^{2}}{m_{h}^{2}}},
\end{align}
where
the coupling constant $\lambda_{haa}$, which is defined by $\mathcal{L}\ni \lambda_{haa}haa$, can be derived from the Higgs potential in the mass basis. 
The numerical results for the branching ratio are obtained by performing a scan analysis presented in Appendix~\ref{ap:mass}. 
As seen from the left panel, the branching ratio ${\rm Br}(h\to aa)$ can be larger than $0.1\%$ if $\theta_{hs}\gtrsim 10^{-2}$. 
In addition, there is a tendency that larger value of $\lambda_{P}$ increases the branching ratios. 
In the right panel, one can see that the magnitude of the branching ratio is almost independent with the ALP mass $m_{a}$ and $c_{h}(=\theta_{ah}m_{h}m_{\Phi}/m_{a}^{2})$  in this calculation since we consider the parameter region where the mixing between $h$ and $a$ is tiny. 
Therefore, the branching ratio which satisfies \Eqs{br} and \eqref{brex} is possible for most of natural parameter space in Fig.~\ref{v}.

We note that in hadron collider, such as (HL-)LHC, one may also have the similar test of the displaced vertex. In some $m_a$ and $L_D$ ranges, that the ALP decays to certain products like a muon pair (See the cases for hidden photon Ref.~\cite{Curtin:2014cca} and usual ALP \cite{Bauer:2017ris}\footnote{See also \cite{Brivio:2017ije} when the ALP does not decay in the detector.}), hadron collider may be more powerful than lepton colliders thanks to the large number of the produced the Higgs bosons. 
In some ranges, it is quite difficult due to the background. 
In addition, the hadron colliders may confirm our scenario by searching for the heavy $s$.
The future lepton collider, on the other hand, may provide a model-independent search and a discrimination of the different dark sector (see Appendix.~\ref{app:1}.)
As we can see in Fig.\ref{fig:br}, most of the natural region can be tested in the future lepton collider, irrelevant to the mass and $L_D$ range. 
 Therefore, a future Higgs factory of a lepton collider may be a nice tool to probe or discriminate a generic dark sector.
In particular, the CP-even ALP has a strong dependence of $L_D$ on $m_a$, which also determines the decay products. The measured $L_D$ and the decay products (and perhaps also the ALP mass from momenta reconstruction) pointing to a similar mass range is a smoking-gun evidence of the CP-even ALP. This also makes it easy to be discriminated from the other light dark sector models.

Before going to the next section, let us also mention a motivation to the parameter region with the decay length not too longer than several kms. The ALP may play the role of a mediator between SM particle and other dark particle, like a DM,
in the early Universe. 
In this case, the ALP may be popularly produced in the early Universe and it should decay much shorter than the age of Universe.
Indeed, if a hidden field is charged under $\U(1)$, e.g. ${\cal L}\supset \F \bar{\Psi^c} \Psi$ with $\Psi$ being a hidden fermion, the coupling between  $a$ and $\Psi$ in the low energy effective theory is not as suppressed as the couplings between the single $a$ and SM particles. This is the desired property to evade various cosmological and astrophysical bounds when a boosted light dark particle affects the ground based experiments~\cite{Jaeckel:2020oet}. 
The ALP itself can be the dominant DM as we will discuss in the next section.

\section{Light CP-even ALP as DM}

\subsection{DM stability and phenomenology}
When the ALP mass is small the decay rate is highly suppressed.\footnote{The possible Planck-suppressed higher dimensional terms may follow the naive dimensional analysis. 
In this case, the possible coupling with $\U(1)$ breaking is suppressed by $\k$ which is negligible. The $\U(1)$ symmetric higher dimensional coupling appears with $|\F|^2$ which preserves the $C_{\rm dark}$ symmetry. 
Since the ALP is the lightest particle under the $C_{\rm dark}$ symmetry, the ALP does not decay via these terms. 
If the higher dimensional term does not follow the analysis,  we may need a gauge symmetry to provide a good quality of the $\U(1)$. A (gauged) $Z_2$ discrete symmetry is enough to suppress the decay rate in the mass range of  $m_a=\KEV-\MEV$ since the Planck-suppressed term starts from dimension 6. 
 }
Thus it is a good candidate of the DM.
To study its nature, let us consider $m_a\ll 2m_e $, in which case the ALP dominantly decays into a photon pair via 
\beq
{\cal L}^{\rm eff} \supset- \frac{a }{\L_\g} F_{\m\n} F^{\m\n}. 
\eeq
$\L_\g$ is a higher dimensional  coupling. 
The decay rate can be given by $\G_{a \to \g\g} \simeq \frac{m_a^3}{4\pi \L_\g^2}$.

The higher dimension term can be found from the Higgs-VEV dependence of the running  behavior of the electric gauge   coupling:
\beq
\laq{RGe}
\frac{a}{\L_g}=   a \theta_{ah}  \times \partial_v \frac{1}{4e[v]^2}. 
\eeq
Note that the $v$ dependence in $e$ may mostly come from  $c, b, t, e, \m,\t, W$ (3 flavor QCD), but it is not very sensitive to $u, d, s$. 
This is because the most hadron masses depend on the QCD scale. 
By taking account of this effect 
we take the decay rate to be c.f. Ref.\, \cite{Liu:2014cma},
\beq
\laq{agg}
\G_{a\to \g\g}= \frac{m_a^3 \theta_{a h}^2}{\pi^5v^2}\times   \{\frac{121}{36864},\frac{1}{4096} \},
\eeq
the former (latter) of which assumes that 
 all the masses of the fundamental fermions (fermions other than $u,d,s$) participate in \Eq{RGe}.

\begin{figure}[!t]
\begin{center}  
   \includegraphics[width=1\textwidth]{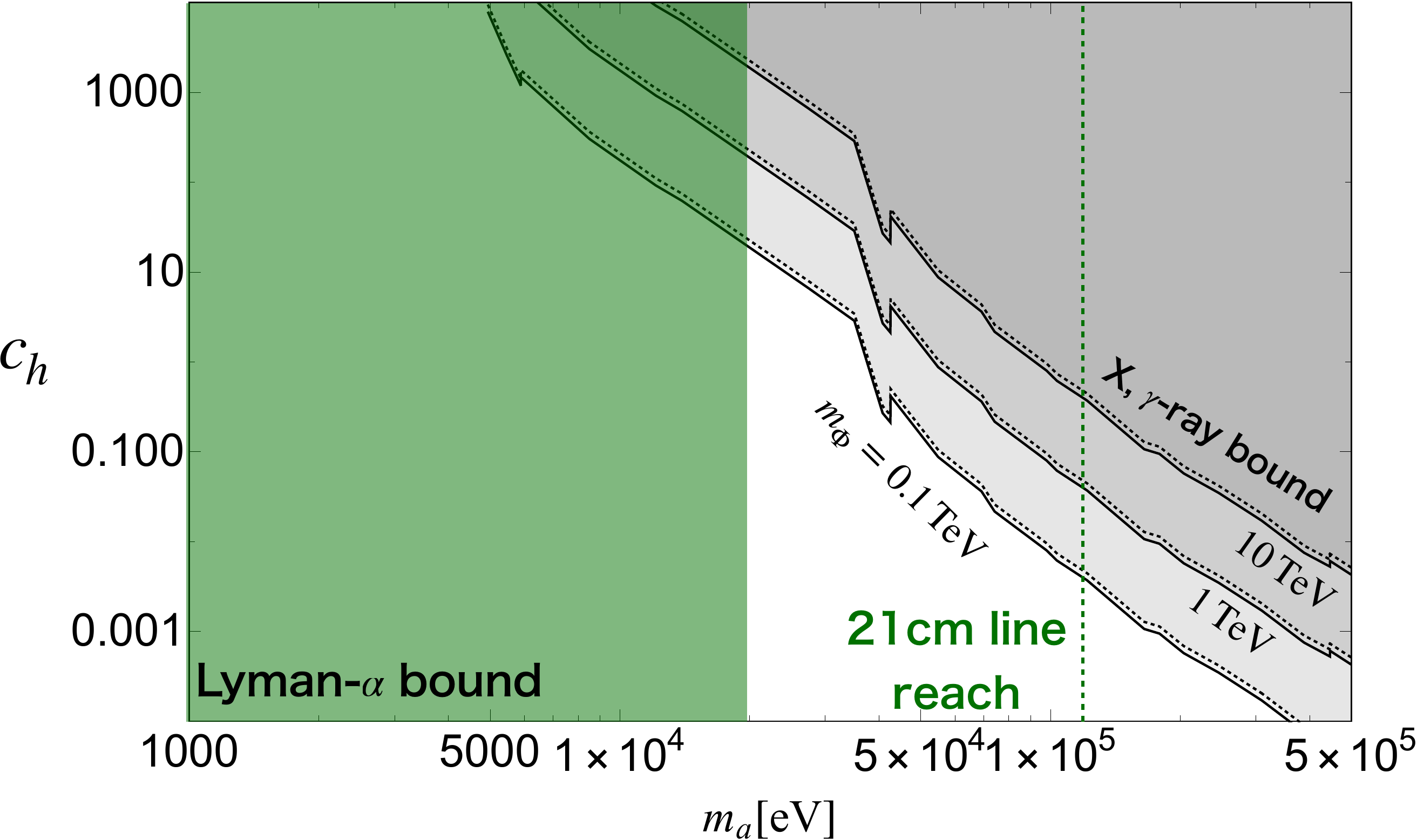}
      \end{center}
\caption{Low mass region of CP even ALP, which is assumed to be the dominant DM. The X $\g$-ray bounds are denoted in the black solid and dotted lines from \Eq{agg} to take account of the hadronic uncertainty, with $m_\F=0.1,1,10\TEV$ from bottom to top. 
The green shaded region (green dashed line) is disfavored (the future reach) by the Lyman-$\a$ bound (21cm-line observation) if the DM is produced thermally. If, on the other hand, non-thermally the bound or reach may not be apply. 
 } \label{fig:DM}
\end{figure}

Since $\theta_{a h}=\O( m_a^2/ m_\F m_h )$, this is extremely suppressed by $m_a^7$. However in the keV-MeV range, the decaying ALP DM can affect the X $\g$-ray  observation. 
By assuming that the CP-even ALP composes the dominant cold DM (see the DM production in the following), we show the parameter region with $m_a<1\MEV$ in 
Fig. \ref{fig:DM}. Also shown is the bound on the thermally produced DM, for which we use $m_a>20\KEV$  (translated from the warm sterile neutrino DM bound $>5.3$keV~\cite{Viel:2005qj,Irsic:2017ixq,Kamada:2019kpe}. If the ALP is a warm DM, the bound should be around this value but slightly larger.).
 This bound depends on the production mechanism, which we assumed the one discussed in the following. However, it can be relaxed with different production mechanisms.  
 The X,$\gamma$-ray bound adopted from \cite{Essig:2013goa} is shown in gray shaded region with several $m_\F=0.1,1,10\TEV$ (The photon flux data are taken from HEAO-1~\cite{Gruber:1999yr}, INTEGRAL~\cite{Bouchet:2008rp}, COMPTEL~\cite{kappadath1998measurement}, EGRET~\cite{Strong:2004de} and FERMI~\cite{Fermi-LAT:2012edv} experiments.). 
 
In the future observation of the 21cm line, which may check the DM with mass $m_a\lesssim 120\KEV$ \cite{Sitwell:2013fpa,Kamada:2019kpe} as well as the future measurement of the $X,\g$ ray via  e.g. ATHENA~\cite{Barret:2018qft}, CTA~\cite{CTAConsortium:2010umy},  
eROSITA~\cite{eROSITA:2012lfj}, Fermi-Lat~\cite{Fermi-LAT:2012edv}, GAMMA-400~\cite{Galper:2012fp,Egorov:2020cmx},  XRISM~\cite{XRISMScienceTeam:2020rvx}, a large parameter region of the DM is testable.\footnote{Let us mention the anthropic selection of the DM mass. 
Conventionally the decaying DM to X,$\g$-rays is difficult to be related to the anthropic principle. 
If there is a bias to larger DM coupling or mass, the anthropic selection is that the parameter is around the critical point where the 
life-time is around the age of the Universe. On the other hand, the bounds from photon observation, which seems not to be anthropic, is so strong that 
the life-time should be much larger than the age of Universe. Thus, the anthropic preferred value is usually excluded. 
In this model, the decay rate is suppressed by $m_a^7$. The anthropic selection may apply to $m_a$ since $m_a\lesssim 0.1 m_a^{\rm crit}$ is enough to evade the X,$\g$-ray bound, where $1/\G_{a\to \g\g}[m_a^{\rm crit}]\sim14{\rm Gyr}$. In this case the DM is preferred to be detected in the near future from the X, $\g$-ray observations. } 
In addition, by using an analytic form of the transfer function~\cite{Konig:2016dzg}  a future reach of the forthcoming Vera Rubin observatory was derived to be $m_a \lesssim \(\frac{g_{\star }}{80}\)^{-1/3}\times 88\KEV$ for a photophilic ALP \cite{Baumholzer:2020hvx} (see also ~\cite{LSSTDarkMatterGroup:2019mwo, Dvorkin:2020xga})\footnote{The reach for our scenario may be slightly higher since the spectrum of the produced ALP is UV dominated due to the dimension 7 operator, while the photophilic ALP is via a dimension 5 operator. On the other hand, by taking account the mass dependence of the transfer function the reach of the 21cm observation may decrease by $\O(1)\%$. }.
 We also remind that the Higgs boson decays invisibly to the DM pair in colliders. 

\subsection{DM production}
Lastly, let us mention some cosmological production mechanisms of the CP-even ALP DM.  
We need the ALP DM not to be thermalized in the early Universe so that it is cold and does not overclose the Universe.
The interaction with a SM fermion, $\p$, by integrating out the Higgs boson is given as
\beq
{\d {\cal L}} =   -\frac{\sqrt{2}{ m_\p}}{\L_H^2 m_h^2} \partial a\partial a  \bar{\psi}\psi.
\eeq
There are also loop suppressed interactions with photons or gluons. 
Those interactions are much stronger than the interaction via the mixing  suppressed by $m_a^2.$
The thermal production rate via $\p\bar \p \to a a $ can be estimated as 
\beq
\G_{\rm th} \sim \frac{1}{4\pi }\frac{ m_\p^2}{\L_H^4 m_h^4} T^7,
\eeq
where $T$ is the cosmic temperature. 
For the ALP not to be thermalized, we need $\Gamma_{\rm th}\lesssim H$ with $H=\sqrt{g_{\star } \pi^2T^4/90M_{\rm pl}^2}$ and $g_{\star}$ being the 
relativistic degrees of freedom. 
We obtain,  
\beq
T \lesssim 7\GEV \(\frac{g_\star}{ 100}\)^{1/10} \(\frac{\L_H}{3\TEV}\)^{4/5} \(\frac{4.18\GEV}{m_\p}\)^{2/5}. 
\eeq
In particular, if the reheating temperature of the Universe is slightly smaller than the r.h.s the ALP can be produced via the same interaction, 
i.e. 
\beq
\Omega_a \sim m_a \times \left.\frac{\G_{\rm th}}{H} \frac{n_\p}{s}\right|_{ T=T_{R}} \times \frac{s_0}{\rho_{c}} \sim  0.35   \frac{m_a}{20\KEV}\(\frac{m_\p}{\GEV}\)^2\(\frac{T_R}{ 2\GEV}\)^5\(\frac{3\TEV}{\L_H}\)^4
\eeq
with $n_\p (\sim T^3/\pi^2), s, s_0 \AND \rho_c$ being the scatterer density, entropy density, present entropy density, and the critical energy density, respectively. 
In this case, the DM has a temperature around $T$. This is hotter than the warm DM of the same mass and obtain a more stringent bound from Lyman-$\alpha$~\cite{Kamada:2019kpe}. 
This reheating may be caused by the inflaton decay, in which case we can obtain the correct baryon asymmetry if the inflaton decays into energetic quarks~\cite{Asaka:2019ocw} which are thermalized after a few flavor oscillations~\cite{Hamada:2018epb, Eijima:2019hey}. In this case, the baryon number needs to be violated 
 via  higher dimensional operators which preserve a baryon parity to stabilize the proton.\footnote{In addition, (another) ALP, or/and right-handed neutrino may be produced from the inflaton direct decays. They propagate over the thermal history. The cascade decaying into the SM particles
 may be detected~\cite{Jaeckel:2020oet,Jaeckel:2021gah, Jaeckel:2021ert}. The spectrum of such dark radiation has a specific feature~\cite{Conlon:2013isa, Jaeckel:2021gah}, by measuring which we can possibly have a strong evidence of the reheating~\cite{Jaeckel:2021gah}.}

We also note a simple scenario that the entropy release during the phase transition dilutes the thermalized DM  by a factor $>\O(10^2)$ e.g. \cite{Yamamoto:1985rd,Lyth:1995ka, Konstandin:2011dr, Iso:2017uuu, Azatov:2021ifm}, but these scenarios may require certain BSM fields coupled to the $\F$ to induce a significant entropy production and a low reheating temperature.

If the ALP is produced non-thermally, on the other hand, e.g. from the decay of inflaton whose mass is comparable or smaller than the reheating temperature~\cite{Moroi:2020has, Moroi:2020bkq}, the DM can be colder than the thermal production. In this case, the DM can be much lighter than keV. This scenario can be also probed via the 21cm line and perhaps via the inflaton search since the inflaton mass is small for having a low reheating temperature. 
For an ALP much lighter than eV we may need both the low reheating temperature and the misalignment production of it~\cite{Preskill:1982cy,Abbott:1982af,Dine:1982ah}. 
In this case a tuning on the ALP potential top or other places to be flat to have a late time onset of the oscillation may be needed to explain the DM abundance~\cite{Nakagawa:2020eeg, Marsh:2019bjr} by a pi-inflation~\cite{Takahashi:2019pqf} (see also the original idea in \cite{Daido:2017wwb}, related low-scale inflation models~\cite{Daido:2017wwb, Daido:2017tbr, Takahashi:2019qmh, Takahashi:2021tff}. ). The tuning can be realized in scenarios with many ALPs by chance or in certain UV models. 

In any case the bounds from fifth force or from star cooling are irrelevant because the extreme suppression of $\theta_{ah} \propto m_a^2/m_h m_\F$ in  the relevant $m_a$ range.

\section{Conclusions}

In this paper, we have studied the phenomenology of a UV complete dark sector model including a massive ALP, by introducing a single dark Higgs field in addition to the SM sector.  
The dark Higgs field naturally couples to the SM Higgs field via the Higgs portal interaction. 
Giving mass to the ALP by renormalizable tiny $\U(1)$ breaking terms without imposing a CP symmetry, we found that the ALP behaves as a CP-even particle in the low-energy effective theory. 
The difference from the ordinary CP even scalar is the naturally small mass and tiny couplings controlled by the ALP mass. 
Since the decay of the Higgs boson is not suppressed, 
the ALP, which is weakly coupled to the light SM particles, can be probed in future Higgs factories with interesting signatures such as the displaced vertices.
When the ALP mass is below the electron mass, it is a good candidate of the DM despite the relatively heavy mass.  
When the ALP is produced thermally, the ALP can be probed in the future 21cm line observation as well as the X,$\g$-ray observations. 
\\

{\it Note added:}  While completing this paper, we found Ref. 
\cite{Bhattacherjee:2021rml} by Bhattacherjee and Matsumoto and Sengupta, who performed a careful and throughout study on the light mediator test via a displaced vertex search in the present or future hadron colliders. 
Indeed, the proposed DELIGHT in \cite{Bhattacherjee:2021rml} provides a great opportunity to discover our CP-even ALP (toghether with the direct search of $s$ in the 100TeV collider). 
In contrast, we discussed the test and discrimination of the light dark sector in the context of future lepton colliders, by assuming a negligible background for a displaced vertex search. In particular, we pointed out a possible measurement of the decay length.

\section*{Acknowledgement}
We thank Keisuke Fujii and Jurina Nakajima for useful discussion on the property of the ILD during the ILC summer camp 2021. 
We are also grateful to Vedran Brdar, who lets us know the future reach of the freeze-in DM from the halo mass function measurement. 
This work was supported supported by JSPS KAKENHI Grant Nos.  20H01894 (K.S.) 20H05851 (W.Y.), 21K20363 (K.S.), and 21K20364 (W.Y.).
\appendix

\section{Discriminating CP-even ALP from hidden photon in the Higgs factory}
\label{app:1}

Instead of providing an explicit breaking term, 
we can gauge the $\U(1)$ in \Sec{1}, and the NGB is eaten by a gauge boson. 
To discuss this possibility the kinetic term of $\F$ and the hidden gauge field should be invariant under the symmetry as 
\beq
\D {\cal L} \supset -\frac{1}{4}F'_{\mu \nu} F'^{\mu\nu}+ |D_\mu \F|^2 
\eeq
where $F'$ is the field strength of the Hidden photon.
$D_\mu \F\equiv (\partial_\mu + i g' A_\mu )\F $ is the covariant derivative of the dark Higgs field, with  $g'$ being the gauge coupling. 
In this case, we can still calculate the invisible decay of the Higgs boson to the longitudinal component of the hidden photon from \eq{hinv} thanks to the equivalence theorem if $g' \ll \sqrt{\l_P}.$ 
This Lagrangian predicts a stable hidden photon.

We can, on the other hand, write the mixing term between hidden and visible photons. 
According to the gauge symmetries, we obtain 
\beq
 \D{\cal L} =  \frac{\e}{2} F'_{\m \n}F_Y^{\m \n} 
\eeq
where $e$ and $F^{\m\n}$ are the gauge coupling and field strength of the $\U(1)_{Y}$ gauge symmetry.
We note that this is the only possible coupling to the SM particles, if we do not introduce additional particles and assume neutrino masses of the dimension 5 $LLHH$ term which together with the Yukawa interactions disallows any flavor specific gauge symmetry and B-L gauge symmetry.

\begin{figure}[!t]
\begin{center}  
   \includegraphics[width=105mm]{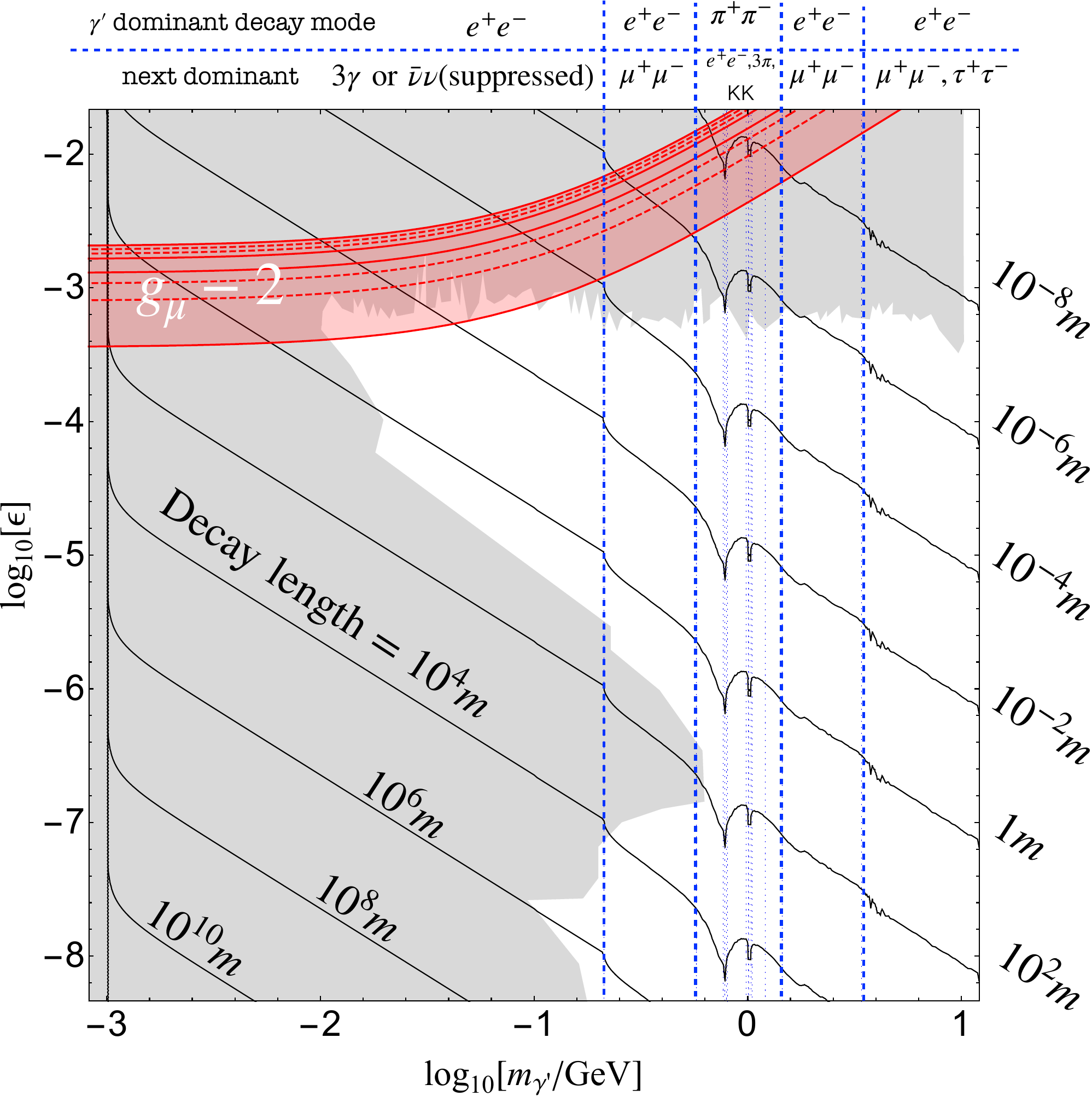}
      \end{center}
\caption{We show the contours of the decay length of the $\g'$ in the exotic Higgs decay, i.e. $h\to \g' \g'$, $\g'\to $SM particles. 
Here we assume the initial Higgs is at rest, which is in good approximation for a $240\GEV \OR 250\GEV$ lepton collider. 
Shown on the top are the dominant and next dominant decay modes  which are of similar 
probability unless otherwise stated. Gray regions are excluded due to various existing constraints adopted from \cite{Liu:2014cma, Filippi:2020kii, Chang:2016ntp}. 
We also show the muon $g-2$ favored region in red. The outer red solid lines denote the $4\sigma$ and the inner $1\sigma$ regions. The $2,3\sigma$ lines are shown in dashed lines. 
 } \label{fig:hidden}
\end{figure}

The proper decay rate of $\g'$ can be expressed as 
\beq
\G_{\rm tot}= \G_{\rm leptonic}+ \G_{\rm hadronic}+\G_{\rm photonic} +\G_{\rm neutrino}
\eeq
The leptonic contribution can be estimated analytically,
\beq
\G_{\rm leptonic} =\sum_{l=e,\m,\t} \frac{1}{3} \a m_{\g'  } \e^2 \sqrt{1- \frac{4m_l^2}{m_{\g'}^2}} (1+2\frac{m_l^2}{m_{\g'}^2})\Theta{[m_{\g'}-2 m_l]}
\eeq
with $\a\approx 1/137$ and $m_{\g'}=\sqrt{2} g' v_\f$ being the hidden photon mass from the Higgs mechanism. 
 The hadronic contribution can be read from the so-called 
 R-ratio of electron-positron annihilations as~\cite{Bjorken:2009mm}
\beq
\G_{\rm hadronic}= \frac{1}{3} \a m_{\g'} \e^2 \frac{\s(e^+ e^- \to {\rm hadrons})}{\s (e^+ e^- \to \m^{+} \m^{-} )}.
\eeq 
We take the R-ratio from PDG~\cite{ParticleDataGroup:2020ssz}.
When the $m_{\g'} <2 m_l$ the radiative induced three body photonic decay $\g' \to 3\g$~\cite{Redondo:2008ec, Alonso-Alvarez:2020cdv},\footnote{The following process is subdominant in the regime the other channel of $\g' \to l\bar l $ is open. We do not consider the scale dependence of photonic decays.  } occurs via
\beq
\G_{\rm photonic}=  \frac{17 \a^4 \e^2}{11664000 \pi^3} \frac{m_{\g'}^9 }{m_l^8}\Theta{[m_l-m_{\g'}]}.
\eeq
Below the mass $m_\g'\ll 5 \KEV$ the process of $\g' \to \nu\nu$ dominates with the rate~\cite{Alonso-Alvarez:2020cdv}
\beq
\G_{\rm neutrino}=  \frac{\a \e^2}{8 \cos^4 \theta_W} \frac{m_{\g'}^5 }{m_Z^4}. 
\eeq
They are extremely suppressed if the leptonic or hadronic channels are open.\footnote{When $m_{\g'}<2m_e$ the dominant one process are the photon and neutrino decays. However, the decay length will be too long to be seen in the collider experiments.  }
In Fig.\ref{fig:hidden}, we display the same figure as Fig.\ref{v} by replacing the ALP to be the hidden photon. 
Also shown in the top of the figure is the dominant and next dominant decay modes~\cite{ParticleDataGroup:2020ssz}. 
The gray regions are excluded due to the ground based experiments as well as the astronomy and cosmology. They are adopted from \cite{Liu:2014cma,Filippi:2020kii}
 and \cite{Chang:2016ntp}. We also denote the favored region of muon $g-2$ anomaly which is recently updated in the Fermilab experiment\,~\cite{Muong-2:2021ojo}.\footnote{See also Ref.\,\cite{Keshavarzi:2019abf}  for the lattice result of the muon $g-2$, which is smaller than that from the $R$-ratio approach. The explanation of the $g-2$ within the SM is an important topic but we may need further checks~\cite{Crivellin:2020zul, Keshavarzi:2020bfy}.}
 \beq
 \D a_\mu= (25.1 \pm 5.9)\times 10^{-10}. 
 \eeq
 by adopting the R-ratio approach~\cite{Roberts:2010cj, Davier:2017zfy,Keshavarzi:2018mgv, Keshavarzi:2019abf, Borsanyi:2020mff, Aoyama:2020ynm, Chao:2021tvp}. The discrepancy is at the $4.2 \sigma$ level. 
 The contribution from the hidden photon can be estimated as ~\cite{Fayet:2007ua, Pospelov:2008zw}
 \beq
\D a^{\g'}_\mu\approx \frac{\e^2 \alpha}{2\pi} \int_{0}^1 dz \frac{2 m_\mu^2 z (1-z)^2}{m_\mu^2 (1-z)^2+m_{\g'}^2 z}
 \eeq
 at the 1loop level. The hidden photon may be one of the simplest ways to induce muon $g-2$ while evading the lepton flavor and CP violation. 
 In the figure we show the $4,3,2,1 \sigma $ favored region from outer to inner in the red band.
 Unfortunately, one can only explain the $g-2$ at the $3-4 \s$ level. That said, with certain modification the $1\s$ explanation can be possible~\cite{Mohlabeng:2019vrz, Tsai:2019buq, Bai:2021nai}.  
 In any case, the favored region, interestingly, can be probed in the Higgs factory by directly seeing the dark photon propagation.
 In the LHC, the decay product of the  electron  may be challenging to be reconstructed, but in the lepton collider such as ILC, it is detectable.

 The decay length and decay modes give an alternative coordinate of the parameter region, from which we can measure the model coupling and mass. 
We note that the dominant and subdominant decay modes are quite different from the CP-even ALP. By measuring both we can discriminate  the CP-even ALP from the hidden photon. In addition, if we can reconstruct the dark particle mass or spin, which should be not impossible in the lepton collider, we can perhaps identify both the model and the parameter region. This measurement is model-independent and should be applied to most of models of light dark sector.
 
\section{ Mass basis for the Higgs bosons}\label{ap:mass}
In this Appendix, we define the mass basis for the Higgs bosons, which is used in numerical evaluation of the branching ratios ${\rm Br}_{h\to aa}$ in the Sec.~\ref{sec:col}.

We parametrize the isospin doublet field $H$ and the complex singlet field $\Phi$ as
\begin{align}
H=
\begin{pmatrix}
G^{+} \\
v+\frac{1}{\sqrt{2}}(\phi_{r}+i\phi_{i})
\end{pmatrix},\quad
\Phi=v_{S}+\frac{1}{\sqrt{2}}(\rho+ia^{\prime}). 
\end{align}
The Higgs potential is defined by Eqs.~\eqref{V} and \eqref{Vexp}, 
here for the shake of simplicity, we replace $m_{\Phi}$ in the soft breaking terms Eq.~\eqref{Vexp} with $\mu_{\Phi}$, which is taken to be input parameters in the numerical calculations of the Higgs branching ratios. 
Three minimisation conditions of the Higgs potential are obtained by the taking the first derivative with respect to $\phi_{r},\rho$ and $a'$, respectively, 
by which $m_{\Phi}$, $m_{H}$ and $c_{4}$ are fixed. 
By the second derivatives of these fields, the mass matrix $\mathcal{M}_{S}^{2}$ in the gauge basis are obtained. 
The off-diagonal components are induced the U(1) breaking term, thus all of $\phi_{r}$, $\rho$ and $a^{'}$ mixes in case of $\kappa \neq 0$. 
We define the physical states in the mass basis as
\begin{align}\label{eq:basis}
\begin{pmatrix}
h \\
s \\
a
\end{pmatrix}=
R_{S}
\begin{pmatrix}
\phi_{r} \\
\rho \\
a^{\prime}
\end{pmatrix},
\end{align}
where the rotation matrix is introduced by
  \begin{eqnarray}  \label{e:Oa}
    R_{S} &=& {R}_{\alpha_3}   {R}_{\alpha_2} {R}_{\alpha_1} \,,
  \end{eqnarray}
with
  \begin{equation}
    \label{e:R}
          {R}_{\alpha_1} = \begin{pmatrix}
                     \cos \alpha_1 & \sin \alpha_1 & 0 \\
            -\sin \alpha_1 & \cos \alpha_1 & 0 \\
            0 & 0 & 1 \end{pmatrix}
          \,, \quad {R}_{\alpha_2} = \begin{pmatrix}
            \cos \alpha_2 & 0 & \sin \alpha_2  \\
            0 & 1 & 0 \\
            -\sin \alpha_2 & 0 & \cos \alpha_2 
          \end{pmatrix}\,,  \quad
          {R}_{\alpha_3} = \begin{pmatrix}
            1 & 0 & 0 \\
            0 & \cos \alpha_3 &  \sin \alpha_3  \\
            0 & -\sin \alpha_3 & \cos \alpha_3 
          \end{pmatrix}.
  \end{equation}
 The mixing angles $\alpha_{1}$, $\alpha_{2}$ and $\alpha_{3}$ corresponds to  $\theta_{hs}$, $\theta_{ah}$ and $\theta_{as}$ in the main text, respectively. 
We identify $h$ is the SM-like Higgs boson with the mass of 125{\rm GeV}, while $s (a)$ is the dark Higgs boson (CP-even ALP). 
Mass eigenvalues for these physical states are derived by applying Eq.~\eqref{eq:basis} to the Higgs potential, 
\begin{align}
{\rm diag}(m_{h}^{2},\;m_{s}^{2},\;m_{a}^{2})=R_{S}\mathcal{M}_{S}^{2}R_{S}^{T} . 
\end{align}
This equation relates the masses and mixing angles to the original potential parameters. 
We replace $\lambda_{P},\; \lambda_{S},\;\lambda_{H},\; c^{H}_{2}$, $c^{\Phi}_{2}$ and $c_{3}$ with the physical parameters. 
In short, we choose the following parameters as a input  in the mass basis:
\begin{align}\label{eq:massinp}
v_{s},\;\;m_{s}^{2},\;\;\mu_{\Phi}^{2},\;\;m_{a}^{2},\;\alpha_{1},\;\;\alpha_{2},\;\;\alpha_{3},\;\;c_{1},\;\;c_{2},\;\;c_{1}^{H},\;\;c_{1}^{\Phi}, \mbox{ (the phase parameters for $c_{i}$, $\tilde{c}^{x}_{i}$)}. 
\end{align}
The mass $m_{h}$ and the electroweak VEV $v$ are taken to be $m_{h}=125.1$GeV, and  $v\simeq174$GeV, respectively. 

In the right panel of Fig.~\ref{fig:1} and Fig.~\ref{fig:br}, we scan 
these parameters in the following ranges\footnote{The linear data distribution are used for the parameters at the vanishing limit of which no symmetry recovers. We also impose a condition to restrict artificial cancellations (see the main text).},
\begin{align}
&m_{a}=10^{[{-2}\;\mathchar`-\; 1]}{\rm GeV},\quad 
m_{s},\;v_{S},\;\mu_{\Phi}=[500{\rm GeV}\;\mathchar`-\; 10{\rm TeV}],\quad   \notag \\
&\alpha_{1}=[0\;\mathchar`-\;2\pi],\quad 
\alpha_{2},\;\alpha_{3}=10^{[{-10}\;\mathchar`-\;-2]}\pi,\quad
\kappa=10^{[{-10}\;\mathchar`-\;-2]}, \notag \\
&c_{1},\;c_{2},\;\tilde{c}_{1}^{H}\;\tilde{c}_{1}^{\Phi}=[0\;\mathchar`-\;1],\quad
\theta_{j},\; \tilde{\theta}^{H}_{k},\;\tilde{\theta}^{\Phi}_{k} =[0\;\mathchar`-\;2\pi],\quad
(j=1\;\mathchar`-\;4,\; k=1,2). 
\end{align}
We take into account theoretical constraints for the scalar coupling constants in the potential with the U(1) symmetry, Eq.~\eqref{V}, namely, perturbativity, $\lambda_{H,P,\Phi}<4\pi$ and the potential to be bounded below  e.g.~Refs.\,~\cite{Barger:2008jx,Lebedev:2012zw,Elias-Miro:2012eoi,Fuyuto:2014yia}, $\lambda_{H,\F} >0$, $ \lambda_{H}\lambda_{\F} >\lambda_{P}^2$ (we take a rather conservative relation to take account of the possible running effect). 
Considering the constraint from measurement of the Higgs signal strength at the LHC ~\cite{Aad:2019mbh, CMS:2020gsy}, we also impose that the Higgs boson coupling with weak gauge bosons does not much deviate from the SM predictions, $g_{hVV}$/$g_{hVV}^{SM}=\cos\alpha_{1}\cos\alpha_{2}\geq 0.995$. 
For the allowed parameter points that pass above the constraints, we further remove some of them, where tuning of parameters could happen. 
Such a criteria are taken as follows; 
(i): the derived dimensionless coefficients $c_{2}^{H}$, $c_{2}^{\Phi}$, $c_{3}^{}$, $c_{4}^{}$
are the same order as the others, i.e., $0< c_{2}^{H}$, $c_{2}^{\Phi}$, $c_{3}^{}$, $c_{4}^{}<1$. 
(ii): the scalar couplings $\lambda_{H,P,\Phi}$ and the dimensionless couplings $c_{i}, \tilde{c}^{x}_{j}$  are not too small, i.e, $\lambda_{H,P,\Phi}>\delta$, 
(iii): The mass of ALP $m^{2}_{a}$ is not too small, compared with the dominant contributions to the mass matrix element in the gauge basis, $(\mathcal{M}^{2}_{S})_{a^{\prime} a^{\prime}}$, i.e., (largest cosine contribution in \Eq{Va}) $\times \delta < m^{2}_{a}$. 
We set the tuning parameter $\delta=5\cdot 10^{-3}$. 
For a sufficient number of data in a limited machine power, we are red to take a  smaller $\d$. In this analysis, whether the vacuum is a global minimum is not checked. However we expect that the result does not change much from the left panel of Fig.~
\ref{fig:1}.

\providecommand{\href}[2]{#2}\begingroup\raggedright

\end{document}